\documentclass[traditabstract]{aa}

\usepackage{lscape}
\usepackage{multirow}
\usepackage{natbib}
\usepackage{graphicx}
\usepackage{float}
\usepackage{amssymb}
\usepackage{amsmath}
\usepackage[dvipsnames]{color} 

\newcommand{\be}{\begin{equation}}
\newcommand{\eeq}{\end{equation}}

\def\ba{\begin{eqnarray}}
\def\ea{\end{eqnarray}}

\def\msol{M_\odot}
\def\mdot{\dot M}

\def\msolyr{M_\odot \rm {yr}^{-1}}
\def\te{T_{\rm eff}}

\def\ltsima{$\; \buildrel < \over \sim \;$}
\def\simlt{\lower.5ex\hbox{\ltsima}}
\def\gtsima{$\; \buildrel > \over \sim \;$}
\def\simgt{\lower.5ex\hbox{\gtsima}}

\newcommand{\cp}{\citep}
\newcommand{\ct}{\citet}

\begin{document}

\title{Self-consistent evolution of accreting low-mass stars and brown dwarfs}

\titlerunning{Self-consistent evolution of accreting low-mass stars and brown dwarfs}

\author{ I. Baraffe \inst{1,2}, V. G. Elbakyan \inst{3}, E. I.~Vorobyov \inst{4,3} and
G. Chabrier \inst{2,1}}
\authorrunning{Baraffe et al.}

\institute{
 University of Exeter, Physics and Astronomy, Stocker Road, EX4 4QL Exeter, UK 
 (\email{i.baraffe@ex.ac.uk})
 \and
Univ Lyon, Ens de Lyon, Univ Lyon1, CNRS, Centre de Recherche Astrophysique de Lyon UMR5574, F-69007, Lyon, France
\and
Research Institute of Physics, Southern Federal University, Rostov-on-Don, 344090, Russia 
\and
Department of Astrophysics, University of Vienna, Vienna, 1180, Austria 
}

\date{}

\abstract{We present  self-consistent calculations coupling numerical hydrodynamics simulations of collapsing pre-stellar cores  and stellar evolution models of accreting objects. We analyse the main impact of consistent accretion history on the evolution and lithium depletion of young low-mass stars and brown dwarfs. These consistent models confirm the generation of a luminosity  spread in Herzsprung-Russell diagrams at ages $\sim$ 1-10 Myr. They also 
confirm that early accretion can produce objects with abnormal Li depletion, as found in a previous study that was based on arbitrary accretion rates. The results strengthen that objects with anomalously high level of Li depletion in young clusters should be extremely rare.
We  also find that early phases of burst accretion can produce coeval models of similar mass with a range of different Li surface abundances, and in particular with Li-excess compared to the predictions of non-accreting counterparts. This result is due to a subtle competition between the effect of burst accretion and its impact on the central stellar temperature, the  growth of the stellar radiative core and the accretion of fresh Li from the accretion disk. Only consistent models could reveal such a subtle combination of effects. This new result could explain the recent, puzzling observations of Li-excess of fast rotators in the young cluster NGC 2264. Present self-consistent accreting models are available in electronic form.}

\keywords{Stars: formation --- Stars: low-mass, brown dwarfs --- Stars: pre-main sequence --- Stars: abundances ---accretion, accretion disks}

\maketitle


\section{Introduction}
\label{sec_intro}

Recent analysis of the effect of  accretion on the structure and evolution of young low-mass stars and brown dwarfs suggests that early accretion history may impact the properties of objects (luminosity, radius and effective temperature) in star-formation regions and young clusters even after several Myr when accretion process has ceased \cp{Baraffe2009, Baraffe2010, Hosokawa2011, Baraffe2012}. This scenario can provide  an explanation for the luminosity spread observed in young clusters, without requiring an age spread \cp{Baraffe2009, Baraffe2012}.  More and more observational evidence suggests that low-mass stars accrete most of their material episodically, through bursts of various intensities. Numerical simulations and analytical studies confirm the existence of such accretion bursts, with different possible driving mechanisms invoked \cp[see][for a comprehensive review]{Audard2014}.
At the very early stages of pre-stellar core collapse,
the accretion bursts, as they release a significant amount of energy,  can have a  profound effect on 
the thermal and chemical evolution of the protostellar disk and the envelope.
They heat up the disk and the inner envelope, raising the gas temperatures by a factor of two or
more \citep{Vor2014} and evaporating CO, CO$_2$, and some other ices 
from  dust grains. These species can linger in the gas phase for a time period much longer than the
burst duration, opening a possibility for the indirect detection of the past bursts
\citep{Lee2007,Visser2012,Vorobyov2013,Jorgensen}.

Accretion bursts can also have a strong impact on the structure of the central object depending on the accretion rate and the amount of accretion energy absorbed by the protostar. Preliminary analysis \cp{Baraffe2009, Baraffe2010} was based on the assumption of ``cold" accretion, assuming that no accretion energy  is absorbed by the protostar.
In a recent work \citep{Baraffe2012}, we propose a so-called ``hybrid" accretion scenario, 
wherein the amount of accreted energy  depends on the accretion rate.  This hybrid scenario  provides a unified picture linking the observed luminosity spread in young clusters and FU Ori eruptions.

Intense accretion bursts could also account for unexpected lithium and beryllium 
depletion in some young cluster members \citep{Baraffe2010,Viallet2012}. Based on arbitrary accretion rates and burst numbers,  \ct{Baraffe2010} find that accretion bursts with rates $>  10^{-4} \msolyr$ could yield significant Li depletion compared to a non accreting model of same age and mass. A recent observational study from \ct{Sergison2013}, based on a careful analysis of star members in the Orion Nebular Cluster and in NGC 2264,  confirm the presence of a dispersion in the strength of Li line, as found in previous observational studies. This dispersion indicates a spread in lithium depletion among the members of these clusters, but the level of depletion remains weak. They also find some evidence for weak Li depletion that correlates with luminosity, which could be either explained by an age effect, or by models of past accretion with rates $< 5 \, 10^{-4} \msolyr$. They however exclude the presence of strongly lithium depleted members in their sample, calling into question the results of \ct{Baraffe2010}. 

As a step forward to  previous models based either on arbitrary or non-consistent accretion rates\footnote{The models of \citet{Baraffe2012} are not fully consistent since the evolutionary calculations of the protostar were based on a post-process treatment using accretion rates predicted by the hydrodynamical simulations of \citet{Vorobyov2010}.  In this work, the protostar feedback on the disk was based on evolutionary models from \citet{Dantona1997}.}, we present in this paper self-consistent numerical simulations fully coupling numerical hydrodynamics models of collapsing pre-stellar cores \citep{Vorobyov2010} and stellar evolution models \citep{Chabrier1997,Baraffe1998}. We analysed the main impact of consistent accretion history, using hybrid and cold accretion assumptions, on the evolution of young low-mass stars and brown dwarfs and on Li depletion. 


\section{Model description}
\label{sec_model}

We consistently modelled the evolution of the central accreting object
and of its circumstellar disk. The model consists of
a forming star, described by a one-dimensional (1D) stellar evolution code (see \S \ref{evolcode}), and a circumstellar disk 
plus infalling envelope described by a two-dimensional (2D) numerical hydrodynamics code (see \S \ref{model}).
Both codes are coupled in real time during the simulations (no postprocessing). 
The input parameter for the stellar evolution code provided by disk modelling is 
the mass accretion rate onto the star $\dot{M}$.
The output of the stellar evolution code is the stellar radius $R_\ast$ and the photospheric 
luminosity $L_{\ast,\rm ph}$, which are used by the disk hydrodynamics simulations 
to calculate 
the total stellar luminosity and the radiation flux reaching the disk surface.
The stellar evolution code is called to update the properties of the protostar 
every 10~yr of the physical time. For comparison, the global hydrodynamical timestep may 
be as small as one month.

The coupled evolutions of the disk and of the protostar are followed up to $\sim$ 1 Myr, when the central object has essentially reached its final mass.  Beyond this point we follow the evolution of the latter with the stellar evolution code alone.  In some cases, the hydrodynamical simulations predict that the central object is still accreting some material from the disk. In those specific cases, we assume that the mass accretion rate predicted by the last hydrodynamical disk simulation declines linearly to zero during another 1.0 Myr. This is certainly ad-hoc, but follows suggestions from observations of  median disk lifetimes $\sim$ 2-3 Myr  with evidence of accretion on the central object still going on \cp{Williams2011}. Since at $\sim$ 1 Myr, the predicted rates from the disk simulations are very small (see \S \ref{sec_results}),  adoption of this arbitrary decay of the rates hardly affects the subsequent evolution of the central object.

\subsection{Stellar evolution calculation}
\label{evolcode}
The stellar evolution models of accreting protostars are based on the input 
physics described in \cite{Chabrier1997} and include accretion processes as described in  
\citet{Baraffe2009} and \citet{Baraffe2012}. 
The accretion rates onto the protostar are derived from the hydrodynamic calculations described 
in Section~\ref{model}.
As in \citet{Baraffe2012}, we assume that an amount $\alpha$ of the accretion energy  
$\epsilon G M_\ast{\dot M}/{R_\ast}$ is absorbed by the protostar, while an amount 
$(1-\alpha$) is radiated away and contributes to the accretion luminosity of the star\footnote{As in \citet{Baraffe2009}, we assume a value $\epsilon$=1/2 characteristic of accretion 
from a thin disk.}. 

 In the present calculations, we consider two scenarios as in \cite{Baraffe2012}: a cold accretion model with
$\alpha=0$, meaning that  essentially all accretion energy is radiated away, and a hybrid accretion
model with $\alpha=0$,  when accretion rates remain smaller than a critical value 
$\dot{M}_{\rm cr}$ and  $\alpha \ne 0$  when $\dot{M} >  \dot{M}_{\rm cr}$.
More specifically, we adopt 
\begin{equation}
\alpha  = \left\{ \begin{array}{ll} 
   0,   &\,\,\, \mbox{if \, $\dot{M}<10^{-5} M_\odot$~yr$^{-1}$ }, \\ 
   0.2,  & \,\,\, \mbox{if \, $\dot{M}\ge 10^{-5} M_\odot$~yr$^{-1}$}. 
   \end{array} 
   \right. 
   \label{function} 
\end{equation}

The justification for a critical value for the accretion rate comes from the analysis of the pressure balance
at the stellar surface and is detailed in Appendix B of \ct{Baraffe2012}. 
The choice for the maximum value of $\alpha$ is based on our previous work showing
that stellar properties do not change significantly as long as $\alpha$ is greater than 
$0.1$--$0.2$ \citep{Baraffe2009,Baraffe2012}. 

Arguments in favour of non zero values for $\alpha$ are provided by the recent radiation hydrodynamics simulations of a protostellar collapse by \ct{Vaytet2013}, following the first and second collapse. They find that the accretion shock at the border of the second (Larson) core is subcritical, implying that essentially all the energy provided by the infalling material is absorbed by the core and not radiated away. 

For the initial seed mass of the protostar,  
corresponding to the second Larson core mass,  we adopt a value of 1.0$~M_{\rm
Jup}$ with an initial radius  $\sim 1.0~R_\odot$. Adopting smaller values for the initial radius would produce even more compact objects \citep[see \S \ref{parameter} and][]{Baraffe2009, Baraffe2012}, while choosing a larger radius ($> 1 R_\odot$)  has no significant impact on the evolution. Indeed, the larger the radius, the smaller the thermal timescale ($\tau_{\rm th} \propto {M^2 \over RL}$) implying that the protostar will rapidly contract and its evolution will be similar to the one starting with 1 $R_\odot$. We do not vary the initial seed mass, as done in previous works \cp{Hosokawa2011, Baraffe2012} on the basis of the calculations of \ct{Vaytet2013}  which point toward the universality of the second core mass of about 1~$M_{\rm Jup}$. 

\subsection{Numerical hydrodynamics calculations}
\label{model}
The evolution of both the disk and envelope is computed using 
the numerical hydrodynamics code described in detail in \citet{Vorobyov2010} and \citet{Vorobyov2013}. 
Here, we briefly recall the main concepts. 
We start our numerical simulations from the gravitational collapse of a starless cloud core, 
continue into the embedded phase of star formation, during which
a star, disk, and envelope are formed, and terminate our simulations after approximately one Myr of
evolution. The protostellar disk, when formed, occupies the inner part of the numerical 
grid, while the collapsing envelope occupies the rest of the grid. 
As a result, the disk is not isolated but is exposed to intense mass loading from the envelope.  
In addition, the mass infall rate onto the disk is not a free parameter of the model 
but is self-consistently determined by the gas dynamics in the envelope.

To avoid too small time steps, we introduce a sink cell near the coordinate origin with a radius of $r_{\rm sc}=5$~AU and impose a free outflow boundary condition so that the matter is allowed to flow out of  the computational domain but is prevented from flowing in.  The sink cell is dynamically inactive: it contributes only to the total gravitational  potential and ensures a smooth behaviour of the gravity force down to the stellar surface. During the early stages of the core collapse, we monitor the gas surface density in  the sink cell and when its value exceeds a critical value for the transition from  isothermal to adiabatic evolution, we introduce a central point-mass object representing the forming protostar. From this time on, the stellar evolution code is used  to calculate the properties of the protostar. We assume that 90\% of the gas that  crosses the sink cell lands onto the protostar. A small fraction of this mass  (a few per cent) remains in the sink cell to guarantee a smooth transition of the gas surface density across the inner boundary.  The other 10\% of the accreted gas is assumed to be carried away with protostellar jets. The use of a sink cell implies that the accretion rate inferred at 5 AU is representative of the protostar's accretion rate. This assumption is a limitation of the models and the reality may be more complex. Additional sources of accretion variability due to thermal
instability \citep{Bell1994} or magneto-rotational instability \citep{Zhu2009}
may be present at sub-AU scales, changing the properties of the accretion rate.

The main physical processes taken into account when computing the evolution of the 
disk and envelope include viscous and shock heating, irradiation by the forming star, 
background irradiation, radiative cooling from the disk surface and self-gravity. Equations and details are provided in 
\citet{Vorobyov2010,Vorobyov2013} and will not be repeated here. The numerical resolution is $256\times 256$ grid points and the numerical procedure to solve the
hydrodynamics equations  is described in detail in \citet{Vorobyov2010}. 
Irradiation from the central star depends on its
luminosity $L_\ast$ which is the sum of the accretion luminosity
 $L_{\rm \ast,accr}=(1-\alpha) \epsilon G M_\ast \dot{M}/
R_\ast$  and
the photospheric luminosity $L_{\rm \ast,ph}$ due to gravitational contraction and eventually deuterium burning
in the protostar interior. The stellar mass $M_\ast$ and accretion rate onto the star $\dot{M}$
are determined self-consistently during numerical simulations, as described in \citet{Vorobyov2013}.

\subsection{Initial setup for the hydrodynamical simulations}
The initial radial profiles of the gas surface density $\Sigma$  and angular velocity $\Omega$ are based on the derivations from  \citet{Basu1997}:
\begin{equation}
\Sigma={r_0 \Sigma_0 \over \sqrt{r^2+r_0^2}}\:,
\label{dens}
\end{equation}
\begin{equation}
\Omega=2\Omega_0 \left( {r_0\over r}\right)^2 \left[\sqrt{1+\left({r\over r_0}\right)^2
} -1\right].
\label{omega}
\end{equation}
Here, $\Omega_0$ and $\Sigma_0$ are the angular velocity and gas surface
density at the centre of the core and $r_0 =\sqrt{A} c_{\rm s}^2/(\pi G \Sigma_0) $
is the radius of the central plateau, where $c_{\rm s}$ is the initial sound speed in the core. 
The gas surface density distribution described by equation~(\ref{dens}) can
be obtained (to within a factor of order unity) by integrating the 
three-dimensional gas density distribution characteristic of 
Bonnor-Ebert spheres with a positive density-perturbation amplitude A.
The value of $A$ is set to 1.2 and the initial gas temperature is set to 10~K.

We have explored a range of initial conditions, characterised by the pre-stellar core mass,
the ratio of the rotational to gravitational energy $\beta$ and the cloud core's outer radius
$r_{\rm out}$. We fix  $r_{\rm out}/r_0=6$ 
in order to generate gravitationally unstable truncated cores of similar form. The adopted values of 
$\beta$  lie within the limits inferred by \citet{Caselli2002} for dense molecular cloud cores. 
To generate the parameters of 
a specific core with a given value of $\beta$, we choose
the outer cloud core radius $r_{\rm out}$ and find $r_0$ using the adopted ratio above-mentioned.
Then, we find the central surface density $\Sigma_0$ from the relation 
$r_0=\sqrt{A}c_{\rm s}^2/(\pi G \Sigma_0)$ and determine the resulting cloud core mass
$M_{\rm core}$ from Equation~(\ref{dens}).



\section{Parameter survey}
\label{parameter}

The range of initial configuration parameters for all simulations are given in Appendix A for the cold accretion (Table \ref{table1}) and hybrid accretion cases (Table \ref{table2}). The motivation for the choice of parameters and number of simulations is to produce a population of young stars and brown dwarfs covering the mass range between $\sim 0.05 \msol$ and $\sim 1.2 \msol$ for which observations in various young clusters are available.
Because of our assumed density structure for the cores and the truncated condition required for collapse, decreasing the core mass implies increasing the gas surface density $\Sigma_0$ and decreasing the radius $r_0$.  Therefore, in order to keep the parameter $\beta$ within the range of observed values, the value of $\Omega_0$ needs to increase if the core mass decreases. Additionally, in order to get gravitationally unstable disks, we choose values of $\beta$ close to the upper observational limit, that is $\sim$ 3\%. These choices of parameters and of initial density structure for the cores yield a correlation of increasing  $\Omega_0$ with decreasing $M_{\rm core}$ (see Tables \ref{table1}-\ref{table2}). We discuss consequences of these choices in \S \ref{hrd}.

Evolution with time of the mass accretion rates onto the protostar since the beginning of the collapse of the pre-stellar cores for all simulations is given in the Appendix Figures \ref{figtmdotcold} and \ref{figtmdothybrid}. To facilitate convergence of the stellar evolution code and avoid too abrupt changes of the accretion rates, actual rates were averaged over the past 50 yr for $\dot M \le 10^{-5} \msolyr$ and 25 yr for $\dot M >10^{-5} \msolyr$.
Differences in the time behaviour of $\dot{M}$ stem from the different properties of protostellar disks formed from the gravitational collapse of pre-stellar cores. The low-$M_{\rm core}$ and low-$\beta$ models produce disks of low mass and size, which are
weakly gravitationally unstable and show no sign of fragmentation, while  
high $M_{\rm core}$ and $\beta$ models form disks that are sufficiently massive and extended
to develop strong gravitational instability and fragmentation. The forming fragments often
migrate onto the star owing to the loss of angular momentum via gravitational interaction with spiral
arms or other fragments in the disk, producing strong accretion bursts similar in magnitude to FU-Orionis-type eruptions \citep[e.g.][]{Vorobyov2010, Baraffe2012}.
In our models, therefore, the stellar accretion and hence the properties of the star
are intimately linked to the parental cores and disk properties.
Depending on whether accretion is cold ($\alpha$=0) or hybrid ($\alpha \ne 0$  for $\dot{M} >  10^{-5} \msolyr$), the impact of accretion on the protostar's structure is different. As already explained in  \citet{Baraffe2009, Baraffe2012}, cold accretion produces objects more compact and thus less luminous at a given age and mass, compared to both hybrid and non accreting counterparts.  Because of the more compact structure of the accreting object in the cold accretion case, the accretion luminosity in this case is larger than in the hybrid case.  
During the major part of the collapse phase, the accretion luminosity radiated away from the protostar into the disk dominates the photospheric luminosity \citep{Elbakyan2016}. 
Therefore, despite the lower photospheric luminosity in the cold case, the total luminosity from the protostar radiated into the disc is larger in the cold case during the major duration of the collapse phase. Because of the feedback effect of the protostar luminosity on the disk,  consistently taken into account, same initial parameters used in cold and hybrid cases will not produce the same accretion history and the same final object. 
We however do not find any obvious trend in the property of accretion and burst intensity resulting from the different feedback effects produced by cold and hybrid accretion respectively. 

\begin{figure}
  \centering
 \includegraphics[width=8cm]{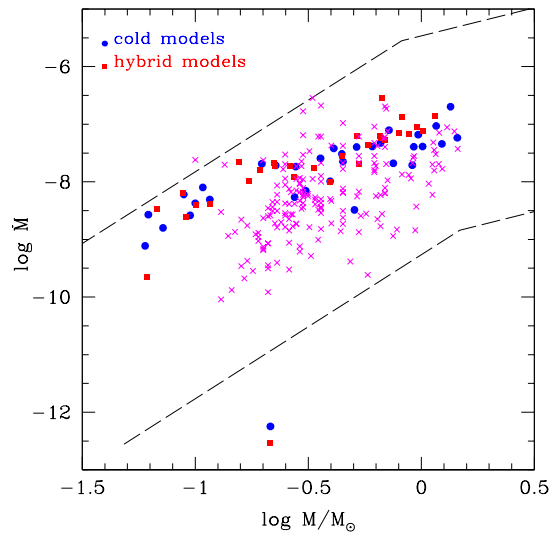}
\vspace{-2cm}
  \caption{Relation between accretion rates derived from the hydrodynamical simulations (at an age of 1 Myr) and  stellar mass for the cold (blue dots) and hybrid (magenta squares) cases. The magenta crosses are taken from the observations of \cite{Venuti2014}. The dashed lines indicate the lower and upper limits observationally derived by \citet{Fang2013}. 
   }
  \label{fig_mmdot}
\end{figure}

A general test for the assumptions used in the hydrodynamical simulations and for the inferred accretion rates during the embedded phases is provided by the work of \citet{Dunham2012}. It shows that the models provide a reasonable match to the observed protostellar luminosity distribution. 
We can also compare present results to rates derived from observations (emission lines, accretion shock emission) in young clusters of age $\simgt$ 1 Myr. This comparison is done in Fig. \ref{fig_mmdot}  with model accretion rates predicted at an age of 1 Myr. 
Observations are based on the recent analysis of the young cluster NGC 2264 ($\sim$3-5 Myr) from \cite{Venuti2014}. We also  
show the  lower and upper limits of the stellar mass-accretion rate relationship derived for Classical T-Tauri Stars (CTTS)  by \cite{Fang2013}. Present comparison between models and observations is  not a proof of validity of the results, given the large uncertainties both in the hydrodynamical models \cite[see][]{Baraffe2012} and  in the determination of  stellar masses, ages and accretion rates from observations. We note that accretion rates for a given mass predicted by the models seem overall larger than the observed rates from \cite{Venuti2014}, particularly for low-mass objects. This could be explained by the fact that the model accretion rates are 
calculated at an age of 1.0~Myr, but the mean age of the NGC 2264 cluster is somewhat older, 3-5 Myr.
In addition, as explained in \S \ref{parameter}, the low-mass models in our sample have $\beta$-values close to the upper observational
limit in order for the disks to be gravitationally unstable, which also results in higher (on average) 
accretion rates.  
We note that previous comparisons of the model and observed accretion rates for 0.5-3.0 Myr-old objects
revealed an overall good agreement \citep{VorBasu2009}. It is therefore reassuring to see that  essentially all predicted rates are within the lower and upper limits of \cite{Fang2013}, and that they follow roughly the observed relations $\mdot \propto M^2$ in the lower mass range ($\log M \simlt -0.5$) and $\mdot \propto M$ in the upper mass range ($\log M \simgt -0.5$) in Fig. \ref{fig_mmdot} \citep[see][]{Fang2013}.

\section{Results}
\label{sec_results}


\subsection{Analysis of the luminosity spread}
\label{hrd}

\begin{figure}
  \centering
 \includegraphics[width=9.5cm, height=18cm]{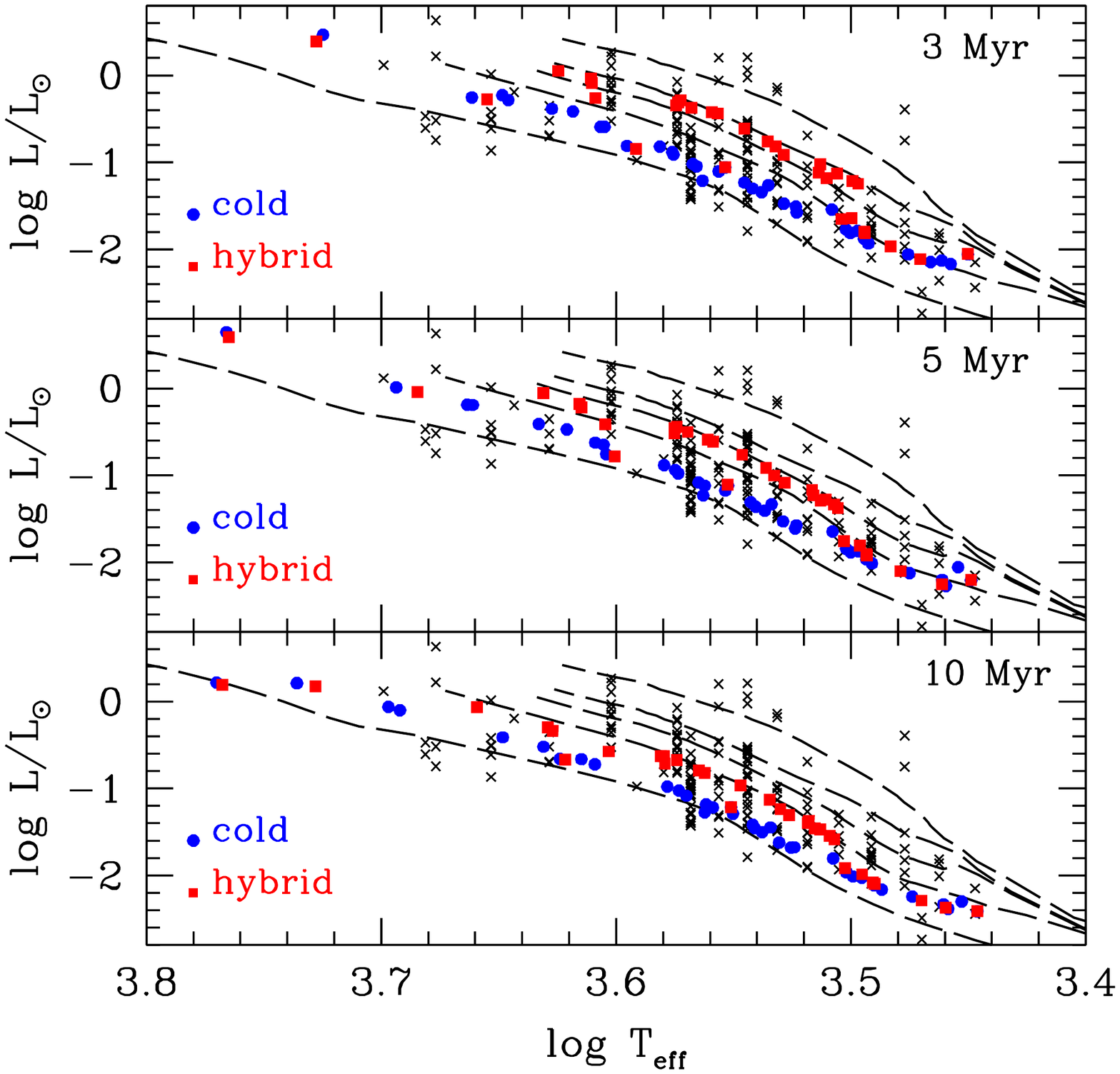}
\vspace{-3cm}
  \caption{Position of cold and hybrid models in a HRD at given ages.  
  The black dashed lines provide isochrones for stellar ages of 1~Myr, 3~Myr, 5~Myr, 10~Myr and 50~Myr, 
  (from top to bottom) derived
from non-accreting stellar evolution models of \citet{Baraffe1998}. The cross symbols are observations from \cite{Bayo2011, Bayo2012} in $\lambda$ Orionis ($\sim$ 5 Myr).}
  \label{fig_hrd}
\end{figure}

In this section we analyse whether the population of synthetic objects resulting from our simulations can naturally produce a luminosity spread in the Herzsprung-Russel diagram (HRD). Figure \ref{fig_hrd} shows the results for cold and hybrid accretion cases for three different ages, namely 3, 5 and 10 Myr. We  exclude models at 1 Myr from the HRD analysis since many of them are still accreting, as illustrated in Fig. \ref{fig_mmdot}. Their accretion luminosity is non negligible compared to the photospheric luminosity, and even exceeds it in a few cases. Observationally, most objects in young clusters with inferred ages of $\sim$ 1 Myr (based on non accreting models, e.g Baraffe et al. 1998) do not show signatures of strong accretion and their luminosity is essentially dominated by the stellar photosphere thermal emission. This suggests that ages inferred for objects as young as $\sim$ 1 Myr are very uncertain and strongly model dependent. 

In order to illustrate the typical observed luminosity spread, we show in Fig.   \ref{fig_hrd} the data  of \citet{Bayo2011, Bayo2012} for the star forming region  $\lambda$ Orionis of typical age $\sim$ 5 Myr. 
More thorough comparison of our models with observations  in young clusters  needs dedicated efforts to obtain observational samples with (i) tested membership, (ii) robust determination of the effective temperature from reliable atmosphere models (presently still lacking) and (iii) case-by-case analysis of very faint objects, which look particularly old, to eliminate objects with edge-on disks \citep[as done in e.g][]{Sergison2013}. 

Two striking features are displayed in Fig. \ref{fig_hrd}. The first is that instead of showing a spread, the models tend to show a bimodal distribution: hybrid models at a given age tend to cluster along the  non-accreting isochrone of same age. The cold accretion models at a given age are systematically fainter and look significantly older (5 to 10 times older) compared to the non-accreting isochrone of same age. The second feature is the lack of luminosity dispersion for $\log \te \simlt 3.5$, that is $\te \simlt$ 3200K. This corresponds to masses $\simlt$ 0.1 $\msol$. Cold and hybrid accretion produce essentially the same objects in the very low mass regime because of the low accretion variability predicted by the simulations and the imposed value of $\dot{M}_{\rm cr}$ which strongly limits the number of bursts during which the accreting object can absorb some amount of the accretion energy (see Fig. \ref{figtmdothybrid}).   

\begin{figure}
  \centering
 \includegraphics[width=10cm, height=16cm]{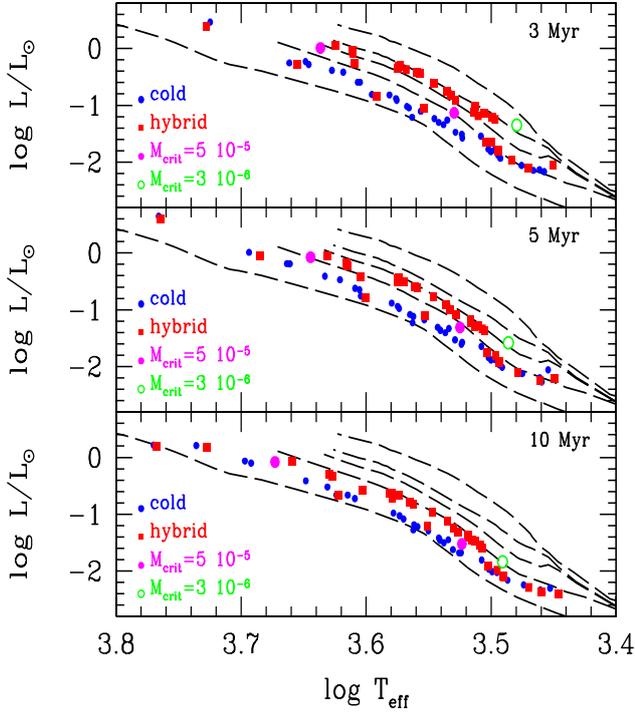}
\vspace{-3cm}
  \caption{Effect of varying the value of $\dot{M}_{\rm cr}$ in the hybrid accretion scenario. The curves are the same as Fig. \ref{fig_hrd}. 
  }
  \label{fig_hrd2}
\end{figure}

One can fill the gaps by varying one parameter in the hybrid scenario, namely the critical accretion rate  $\dot{M}_{\rm cr}$ above which accretion switches from cold to hot. We recall that fiducial values in our hybrid scenario are $\dot{M}_{\rm cr} = 10^{-5}$ and $\alpha$=0.20.
To test the sensitivity of the results to this parameter, we have rerun  two fully coupled sequences with parameters of models 10 and 28, respectively, but with $\dot{M}_{\rm cr} = 5 \times 10^{-5}$ and one sequence with parameters of model 6 but with $\dot{M}_{\rm cr} = 3 \times 10^{-6}$ (see Table \ref{table2} for the corresponding parameters). Increasing the value of $\dot{M}_{\rm cr}$ limits the heating of the protostar due to the absorption of accretion energy during bursts, and thus produces an object with a structure intermediate between the ones produced by cold and fiducial hybrid scenarios, respectively. This helps filling the gap in the HRD for objects with $\log \te \simgt 3.5$ as illustrated in Fig. \ref{fig_hrd2}. We note that
decreasing the value  of $\alpha$ would have the same effect. On the opposite, decreasing the value of  $\dot{M}_{\rm cr}$ allows for more absorption of accretion energy, and thus more heating  of the structure of the accreting object at the low end of the mass distribution, limiting the effect of mass accretion and producing less compact objects.  This helps producing a luminosity spread for the lowest mass objects with $\log \te \simlt 3.5$ as illustrated in Fig. \ref{fig_hrd2}. Justifications for varying these two  parameters $\alpha$ and $\dot{M}_{\rm cr}$ are discussed in \S \ref{discussion}.

As mentioned in \S \ref{parameter}, our choice of parameters for the initial core properties favours fast rotating low mass cores.  More slowly rotating cores, however, certainly exist. They will form low mass objects with small stable disks, with steady instead of burst-like accretion. If, under such conditions, the accretion timescale, $\tau_{\rm acc} = {M \over \dot M}$, remains smaller than the thermal timescale,  $\tau_{\rm th}$, of the protostar, these objects will be representative of the cold accretion case and explain the faintest observed young cluster members \citep[see Fig. \ref{fig_hrd} and][]{Baraffe2009, Baraffe2012}. In the opposite case, $\tau_{\rm acc} >  \tau_{\rm th}$, the objects will have a larger radius, thus a larger luminosity. Therefore, this population of small slowly rotating cores, not explored in the present study, will help resolving the problem of reduced luminosity spread in the low-mass domain.

\subsection{Lithium depletion}
\label{lithium}
\begin{figure*}[h]
  \centering
 \includegraphics[scale=0.42]{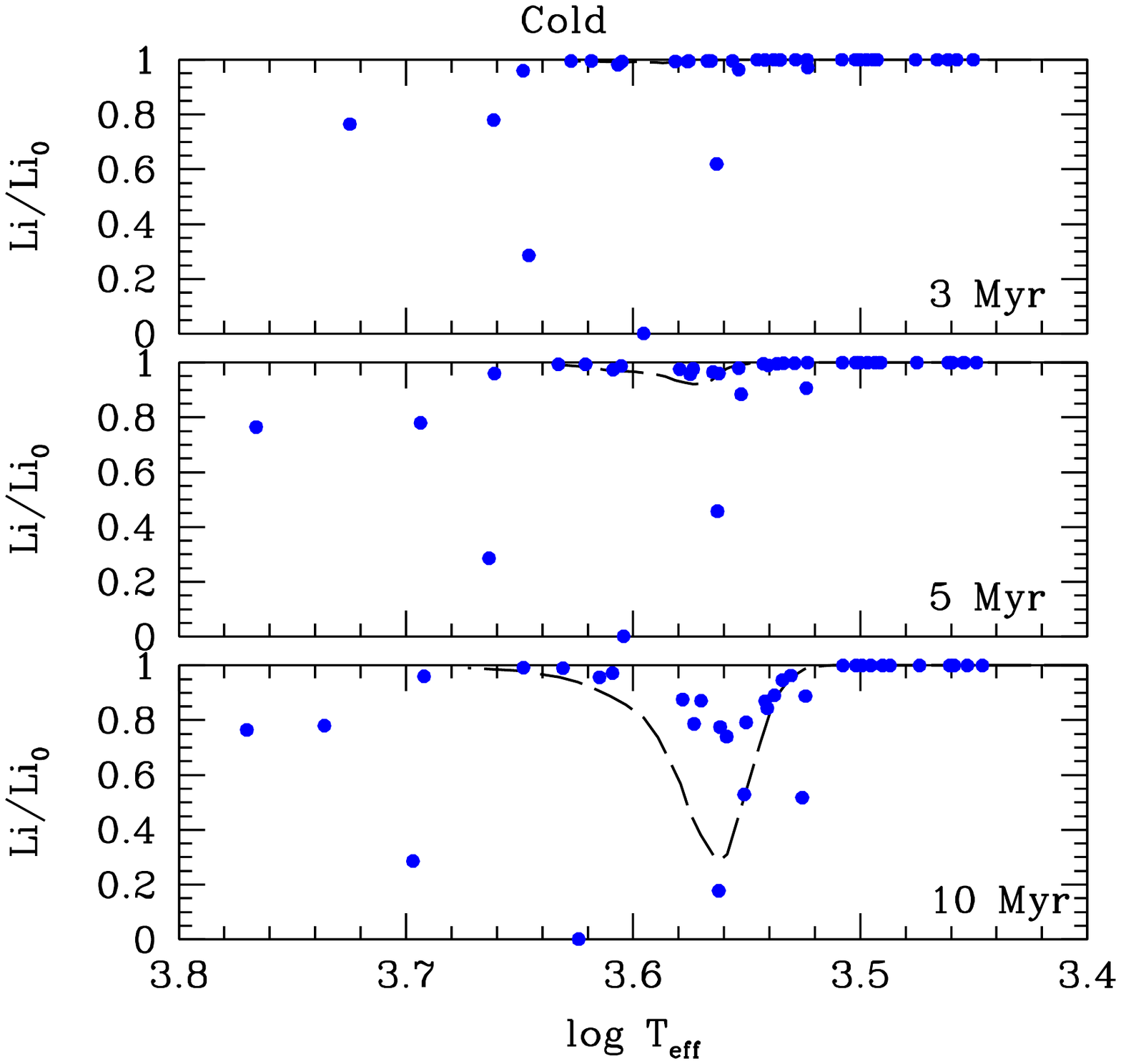}
   \includegraphics[scale=0.42]{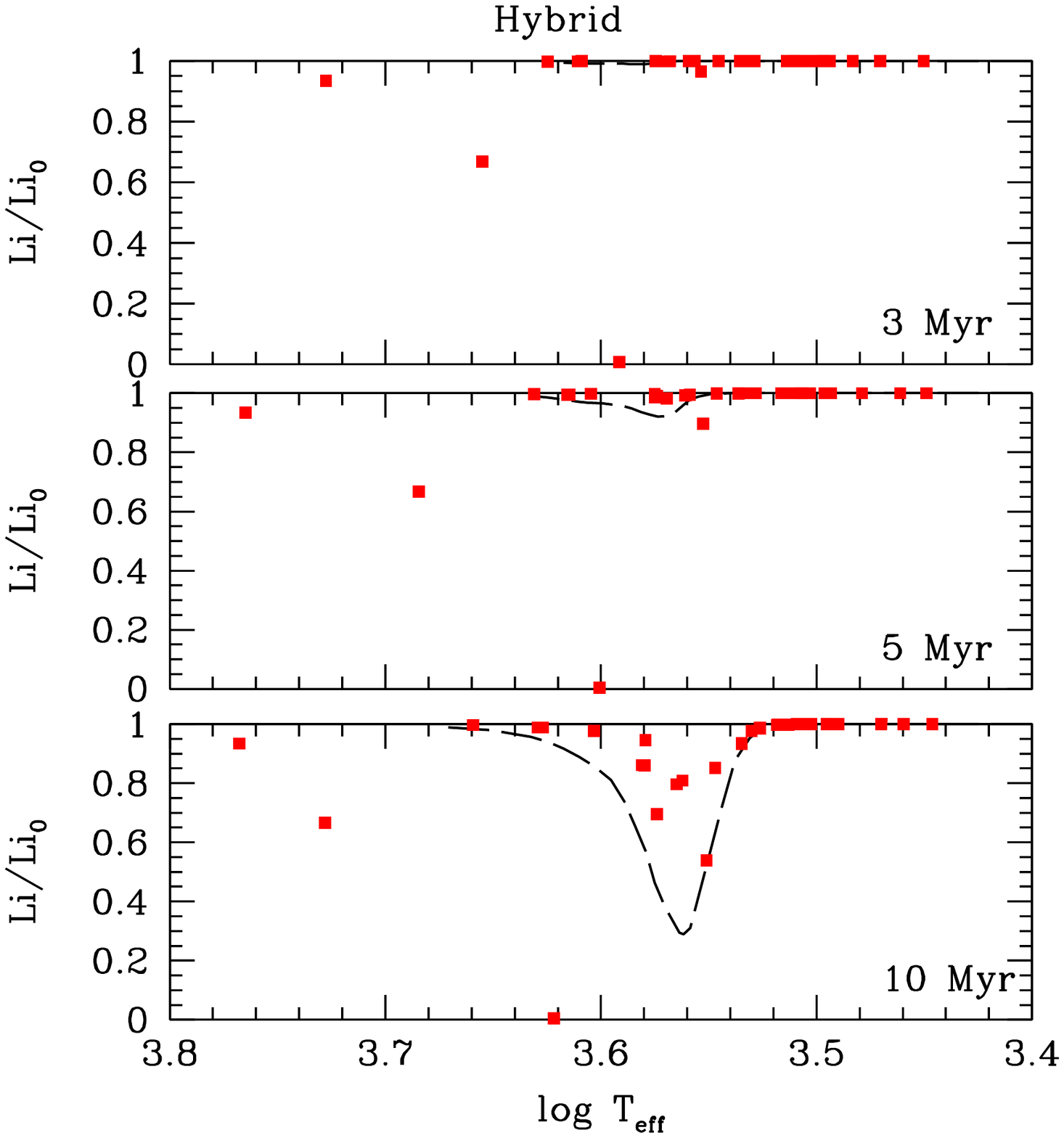}
\vspace{-2cm}
  \caption{Surface lithium abundance (in units of the initial mass fraction Li=10$^{-9}$) as a function of effective temperature in accreting models under the cold (left panels) and hybrid (right panels) accretion scenarios. The dashed curves are the predictions from non-accreting stellar evolution models of \citet{Baraffe1998}. The latter is not visible in the 3 Myr panels since non-accreting models do not predict any depletion at this age and thus Li/Li$_0$=1 for all models.}
  \label{fig_teli}
\end{figure*}

\begin{figure*}[h]
  \centering
 \includegraphics[scale=0.42]{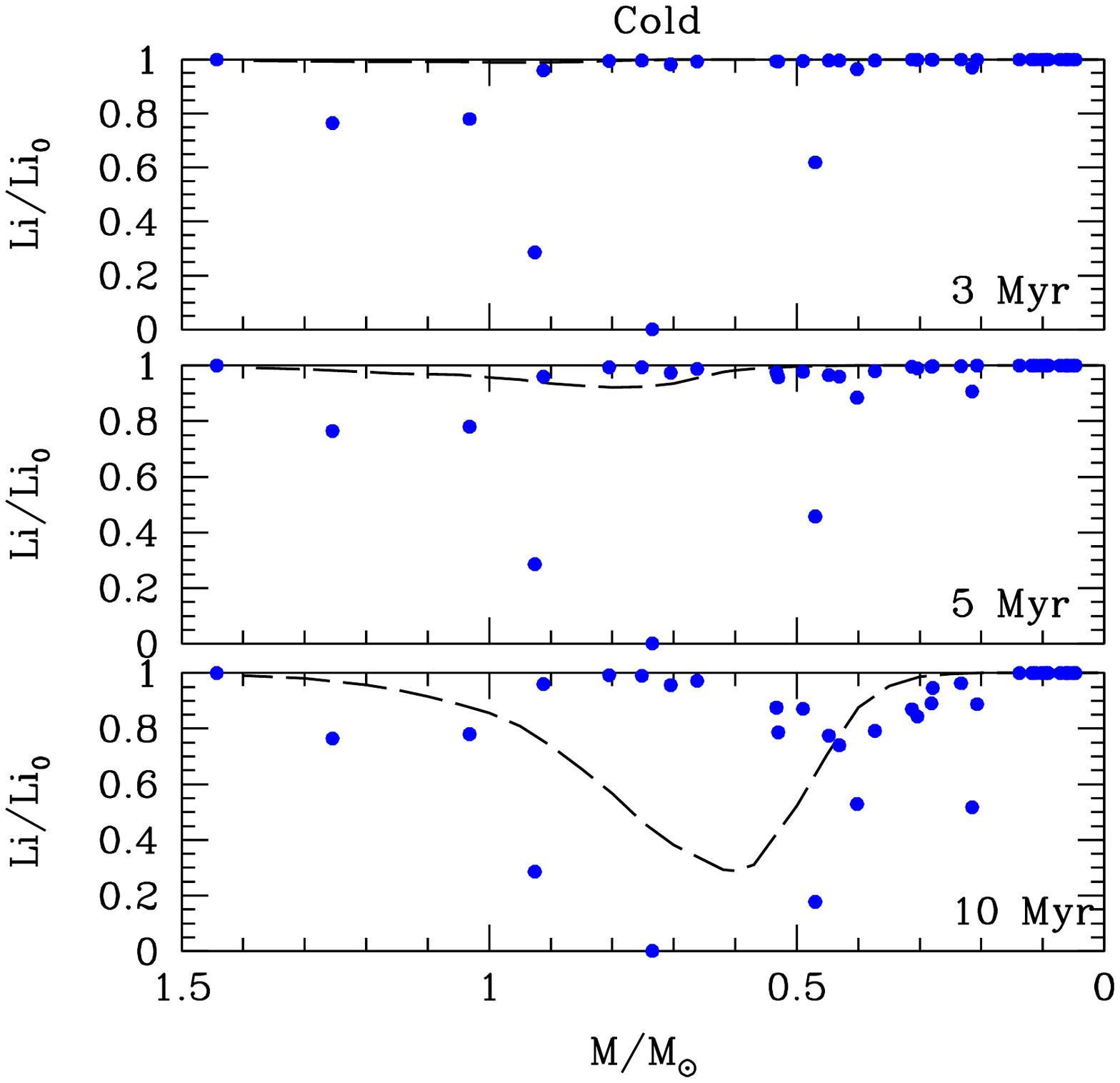}
  \includegraphics[scale=0.42]{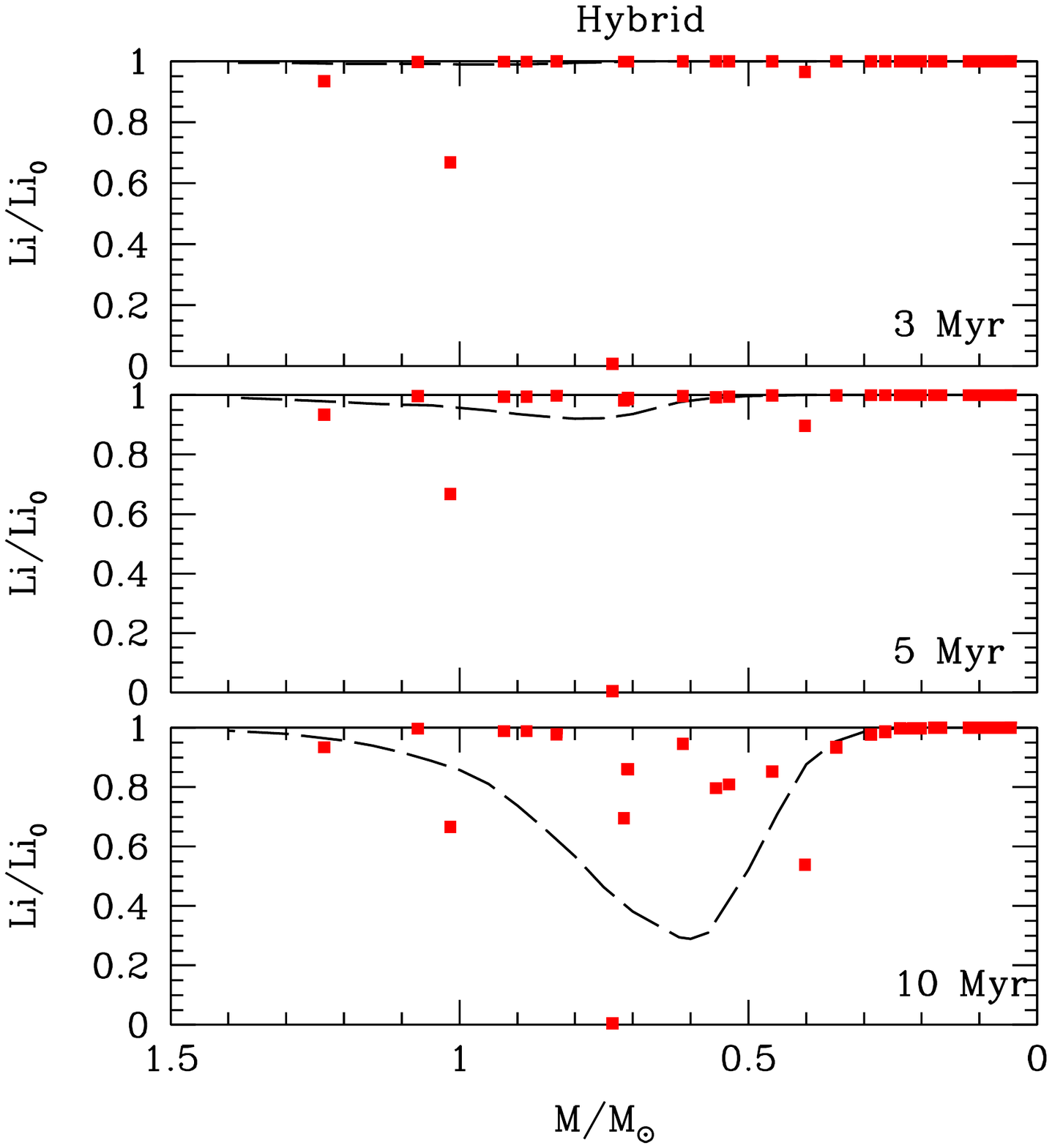}
\vspace{-2cm}
  \caption{Same as Fig. \ref{fig_teli} but as a function of mass.}
  \label{fig_mli}
\end{figure*}

Present consistent calculations enable to re-analyse on more solid grounds the results of \citet{Baraffe2010} about lithium depletion. The possibility of having strong lithium  depletion due to early accretion needs to be confirmed based on more consistent model assumptions since this prediction was called into question by the observations of \ct{Sergison2013}. Results are displayed in Figs. \ref{fig_teli} and \ref{fig_mli} which show the surface abundance of lithium as a function of effective temperatures and stellar mass, respectively, for three different ages.  We summarise the main results below.

\begin{figure*}
  \centering
 \includegraphics[scale=0.50]{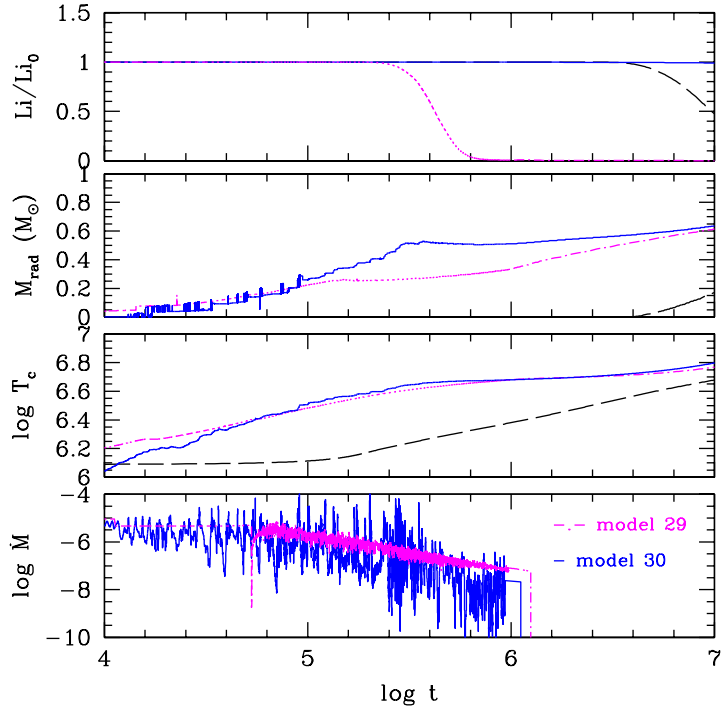}
\vspace{-2cm}
  \caption{Evolution of various quantities as a function of time (in yr) for model 29 (dash-dotted magenta curve)  with final mass 0.735 $\msol$ and model 30 (solid blue curve) with final mass 0.752 $\msol$ under the assumption of cold accretion. Li/Li$_0$ is the surface abundance of lithium in units of the initial mass fraction. $M_{\rm rad}$ is the size of the radiative core in $\msol$. $T_{\rm c}$ is the central temperature and $\dot M$ (in $\msolyr$) is the accretion rate predicted by the hydrodynamical simulations. The black long-dashed curves are the predictions from a non-accreting stellar evolution models with mass 0.75 $\msol$ from \citet{Baraffe1998}.}
  \label{fig_teli_mod29-30}
\end{figure*}

\begin{itemize}
\item{(a)} Both cold and hybrid accretion scenarios confirm the possibility of accelerating lithium depletion at ages of a few Myr compared to non accreting models. As explained in \citet{Baraffe2010}, this stems from the more compact and hotter structure produced by mass accretion. Hybrid accretion, as expected, tends to moderate this effect, producing less anomalously Li-depleted objects compared to non-accreting models. 
\item{(b)} Among all calculated models, we  find only one extremely depleted case compared to the non accreting counterpart  (Li/Li$_0 < $1\% at $t \sim$ 1 Myr). This is model 29 in the cold case (final mass 0.735 $\msol$)\footnote{Model 29 is similar to model 25 in the hybrid case since accretion rates never exceed $\dot{M}_{\rm cr}$ (see Figs. \ref{figtmdotcold}-\ref{figtmdothybrid}). Accretion thus remains cold during the whole evolution.}. Interestingly enough, this sequence is not characterised by strong accretion bursts, as shown in the lower panel of Fig. \ref{fig_teli_mod29-30} (see point (c) below). 
\item{(c)} \citet{Baraffe2010} obtained extremely high level of depletion when the accreting object undergoes from the beginning of its evolution consecutive violent bursts exceeding $\sim$ $10^{-4} \msolyr$. The  hydrodynamical disk evolution models predict the existence of strong bursts in excess of $10^{-4} \msolyr$.  
However, models based on consistent accretion rates show no clear correlation between the existence of such strong bursts  and anomalous level of Li depletion. For example, model 25 (final mass 0.53 $\msol$) with cold accretion displays several extreme bursts  up to $\sim$ $10^{-3} \msolyr$ (see Fig. \ref{figtmdotcold}) but shows no abnormally high Li depletion at an age of a few Myr (Li/Li$_0$ = 0.993 at 3 Myr). On the other hand, model 33 (final mass 0.926 $\msol$) with cold accretion shows similar intense bursts and has anomalously depleted its Li by more than a factor three at 3 Myr ( Li/Li$_0$ = 0.286 at 3 Myr). In the case of hybrid accretion, as mentioned in (a) above, models with the strongest bursts show no anomalous Li depletion.


\item{(d)} As a new result, Figs. \ref{fig_teli} and \ref{fig_mli} show accreting models with higher surface Li abundance than non-accreting counterparts at same age. Those are models in the mass range $\sim$ 0.5 - 1.1 $\msol$ (see Fig. \ref{fig_mli}) which  develop a radiative core during their evolution. This unexpected result is due to a subtle competition between different effects: (i) the more compact structure and the consequent increase of central temperature due to mass accretion, (ii) the growth of the radiative core which strongly depends on the central temperature and density \citep[see e.g][]{Baraffe2010} and (iii) the accretion of fresh lithium from the accretion disk. 

\item{(e)} The results show that all models with higher Li depletion compared to their non accreting counter-part appear to be fainter in a HRD and thus look  older than the non-accreting models. But we stress that there is no systematic correlation between the position in the HRD and the level of lithium depletion, meaning that all objects that look "older" are not necessarily anomalously depleted in Li. As an example, the cold accretion models 22 (final mass 0.448) and  23 (final mass 0.470) have the same position in the HRD, lying at 5 Myr on the 50 Myr non-accreting isochrone. But model 22 has a Li surface abundance depleted by a factor $\sim$ 2 whereas model 23  shows essentially no Li depletion at 5 Myr. 
\end{itemize}

This ensemble of results shows that there is no simple rule or trend linking the intensity of an accretion burst, the level of  Li depletion and the position in the HRD. The evolution of Li is very sensitive to  the strength of  an accretion burst and whether the burst happens when a radiative core is already developed. But it also depends on how rapidly the convective envelope shrinks  and on the relative amount of fresh Li that will still be accreted compared to the abundance of Li in the shrinking convective envelope. Such high sensitivity of results is illustrated in Fig. \ref{fig_teli_mod29-30}:  at time t $\sim$ 0.1 Myr model 30 undergoes several strong accretion bursts which rapidly heat up the central stellar region and cause a rapid growth of the radiative core before a significant amount of Li could be depleted. The additional accretion of fresh Li in a convective envelope which is smaller than that of model 29 or of the non accreting counterpart,  yields a young object with a Li abundance in the convective envelope that is essentially the same as the initial Li abundance in the collapsing pre-stellar core.

\section{Discussion and conclusion}
\label{discussion}

The self-consistent evolutionary models of accreting objects presented in this work\footnote{Models are available at:

http://emps.exeter.ac.uk/physics-astronomy/staff/ib233 

or http://perso.ens-lyon.fr/isabelle.baraffe/accretion\_models} confirm the prediction of a spread in luminosity in HRDs for a coeval population of young low-mass stars and brown dwarfs and the possibility to accelerate lithium depletion.  
Admittedly,  the spread in the HRD predicted by present models is not fully satisfactory. It resembles a bimodal distribution rather than a true spread in luminosity.  Cold accretion models are systematically fainter than the non-accreting isochrone of same age while hybrid accretion models lie close to it.  Limiting the values of free parameters in the present self-consistent calculations with   $\alpha$ fixed either to zero (cold) or to 0.2 (hybrid),  and a fixed value for $\dot{M}_{\rm cr}$ is too simplistic and only provides limiting cases. The fact that, within this restricted choice of parameters, the models produce a limited luminosity spread could be solved by using different values for $\alpha$ and $\dot{M}_{\rm cr}$.

Determination of the value of $\alpha$ characterising the amount of energy transferred to a central object by material accreted from a disk is a formidable radiative hydrodynamical problem that is far from being solved. In the context of young low mass objects, whether accretion is cold or hot is an open debate. Currently, neither models nor observations can solve this question \citep[see e.g.][]{Hartmann2011, Baraffe2012}. 

Assuming that accreted material does reach the surface of the central object with some amount of  internal energy, justifications for varying   $\dot{M}_{\rm cr}$, which delineates the regimes between cold and hot accretion, can be found in the recent multi-dimensional simulations of accretion onto young stars by \cite{Geroux2016}. First, these multi-D simulations provide a test for the assumptions used in 1D evolutionary calculations of accreting objects as presented in this work. They are reassuring regarding the validity of {\it qualitative} results based on 1D accreting models. One interesting finding is that the treatment of hot accretion (for $\alpha \simgt$  0.1) is likely more realistic if based on an outer accretion boundary condition at the stellar surface rather than assuming redistribution of accretion energy deep in the interior, as presently assumed in our 1D evolutionary calculations. \cite{Geroux2016} show that for a given accretion rate and a given value of $\alpha$, the later assumption may overestimate the effect on the structure, in terms of radius expansion of the object. But the important conclusion is that the effects described by the multi-D simulations for a given $\alpha$ can be mimicked by a treatment assuming
redistribution of energy, provided a smaller value of $\alpha$ is used. 

The multi-D simulations also show that there is a change in behaviour at a given value of $\alpha$ in the redistribution of accreted mass and energy in the stellar interior and in the effect of accretion on the stellar structure. The threshold value of $\alpha$ characterising this change in behaviour depends on the mass accretion rate and on the difference between the entropy of the accreted material and the bulk entropy in the convective envelope of the accreting object. It thus highly depends on the structure of the accreting object, and thus on its mass.
This  suggests that the effect of hot accretion, which essentially expands the object and compensates for the effect of mass accretion (which makes the object more compact and fainter)  cannot be  mimicked by one single value of $\alpha$ and a fixed value of $\dot{M}_{\rm cr}$, but more likely by values that vary depending on the accretion rate {\it and} the mass (or the interior bulk entropy) of the accreting object. Work is in progress to explore the implications of the multi-D simulation results. The goal is to  examine whether the use of an accretion boundary in the 1D evolutionary calculations can more naturally produce a luminosity spread than  using the assumption of redistribution of accreted energy in the stellar interior and varying both the values of $\alpha$ and $\dot{M}_{\rm cr}$.

The analysis of Li depletion suggests that extreme Li depletion due to early accretion should  be a very rare event. Among $\sim$ 60 models calculated under various assumptions and parameters, we only find one model which displays almost complete Li depletion. This is not in contradiction with the results of \cite{Sergison2013} who find no strong Li depleted objects among $\sim$ 168 targets in NGC 2264 and the Orion Nebular Cluster.  Interestingly enough, \cite{Bouvier2016} recently re-analysed Li abundance in NGC 2264 members with a sample of $\sim$ 200 stars. They report one object with low lithium equivalent width but with radial velocity consistent with the cluster's average and with rotational velocity $v \sin i$ consistent with a young age. This source was rejected from the \cite{Bouvier2016} analysis but we urge its follow-up with further membership analysis, since confirmation of its membership could provide a genuine signature of our accretion scenario. Our results also show that models with high Li depletion compared to non-accreting models are also fainter in a HRD. But we stress that all objects that look older in a young cluster should not necessarily show a lower level of Li depletion. 

The interesting result about Li depletion in accreting models is the prediction of Li-excess compared to surface Li abundance in
the non-accreting counterparts. This excess is due to a subtle competition between the effect of burst accretion and its impact on the central temperature, the  growth of the radiative core and the accretion of fresh Li. Only consistent models could reveal such a subtle combination of effects. In their analysis of NGC 2264, \cite{Bouvier2016} suggest a connection between lithium abundance and rotation, with  Li-excess in fast rotators compared to slow ones. This discovery is at odd with  current understanding of rotation in stars with rotational mixing enhancing Li depletion. Our results suggest a possible explanation for this puzzle. Strong accretion bursts that occur when the central object develops a radiative core could both explain the higher spin of the star, as a result of angular momentum accretion, and its slower Li depletion due to rapid shrinkage of the convective envelope and accretion of fresh Li.  This idea is attractive and deserves further exploration by both observations and models.

To finish, both hydrodynamical and evolutionary models presented in this work contain many uncertainties, part of them already discussed in \cite{Baraffe2012}. Recent multi-D simulations of realistic stellar structures as presented in \cite{Geroux2016} provide some preliminary tests of major assumptions used in 1D stellar evolution models of accreting objects and give some more confidence in their qualitative predictions. Much more work needs to be done before we can believe in quantitative results and build realistic synthetic populations of young objects accounting for their early accretion history to compare with observations. We believe this work provides one step forward in this direction. But guidance from observations are key to advance on the front of model developments. We suggest within this context more observational efforts devoted to characterising the properties of young objects undergoing accretion bursts such as FUors and EXors. Lastly, the standard practice of rejecting outliers with abnormal Li depletion in young clusters may be a lost opportunity to find a genuine confirmation of our accretion scenario. We strongly advise for a follow-up studies of such outliers.

\acknowledgements
We thanks Amelia Bayo, Min Fang and Laura Venuti for providing their data.
This project was partly supported by the European Research Council through grants ERC-AdG No. 320478-TOFU
and No. 247060-PEPS, by the Russian
Ministry of Education and Science Grant 3.961.2014/K
and by the Austrian Science Fund (FWF) under research grant I2549-N27. V.G.E. acknowledges Southern Federal University Development Program for financial support.
The simulations were performed on the Vienna
Scientific Cluster (VSC-2), on the Shared Hierarchical Academic Research Computing Network 
(SHARCNET), on the Atlantic Computational Excellence Network (ACEnet) and on the University of Exeter Supercomputer, a DiRAC Facility jointly funded by STFC, the Large Facilities Capital Fund of BIS and the University of Exeter.

\appendix
\section{Initial model data}
\begin{figure*}[h]
  \centering
  \includegraphics[width=17cm]{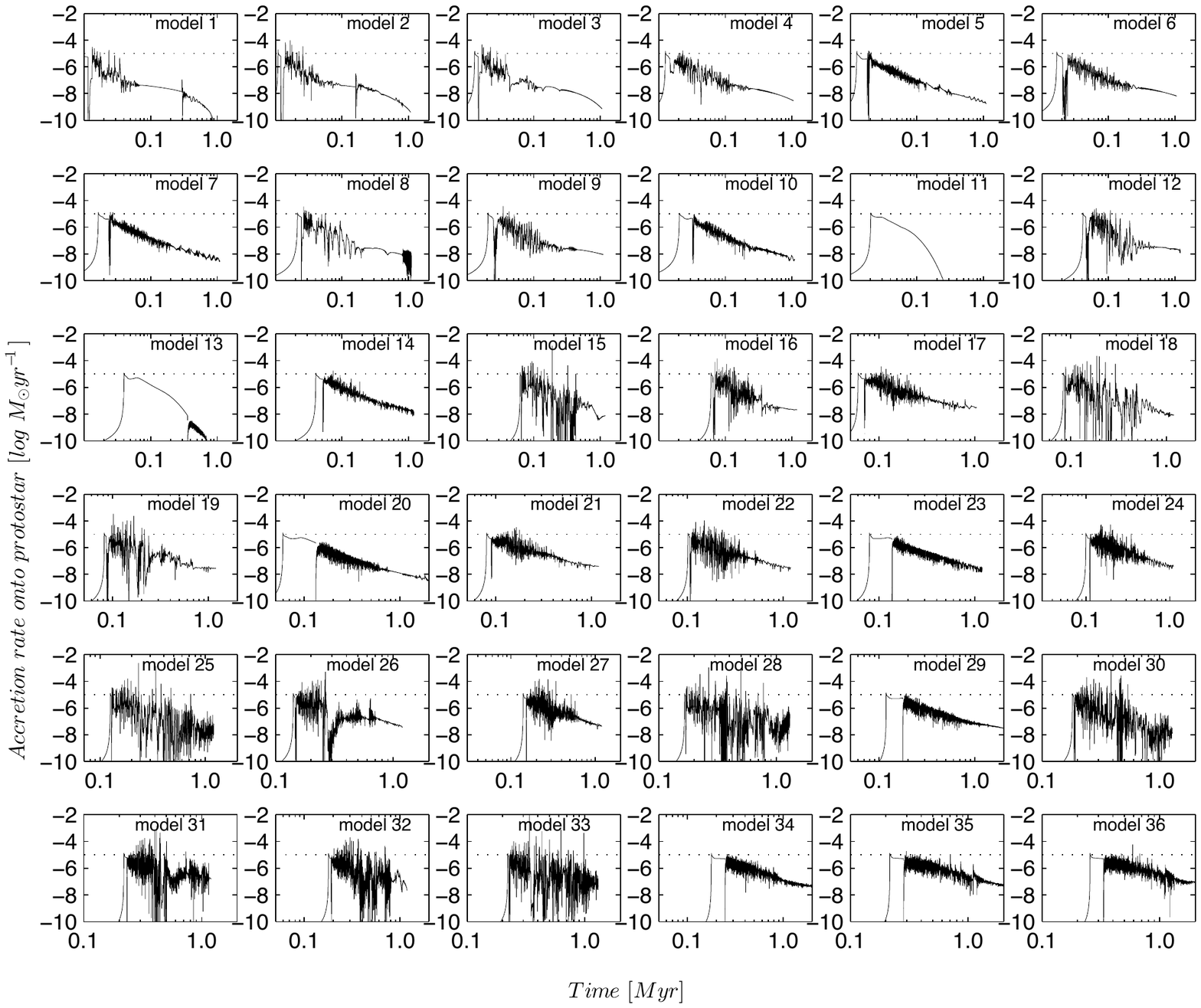}
  \caption{Mass accretion rates vs. time for the 36 models calculated for cold accretion. 
  The horizontal dotted lines mark the critical value for the transition from  cold to hot accretion.
  }
  \label{figtmdotcold}
\end{figure*}

\begin{figure*}[h]
  \centering
  \includegraphics[width=17cm]{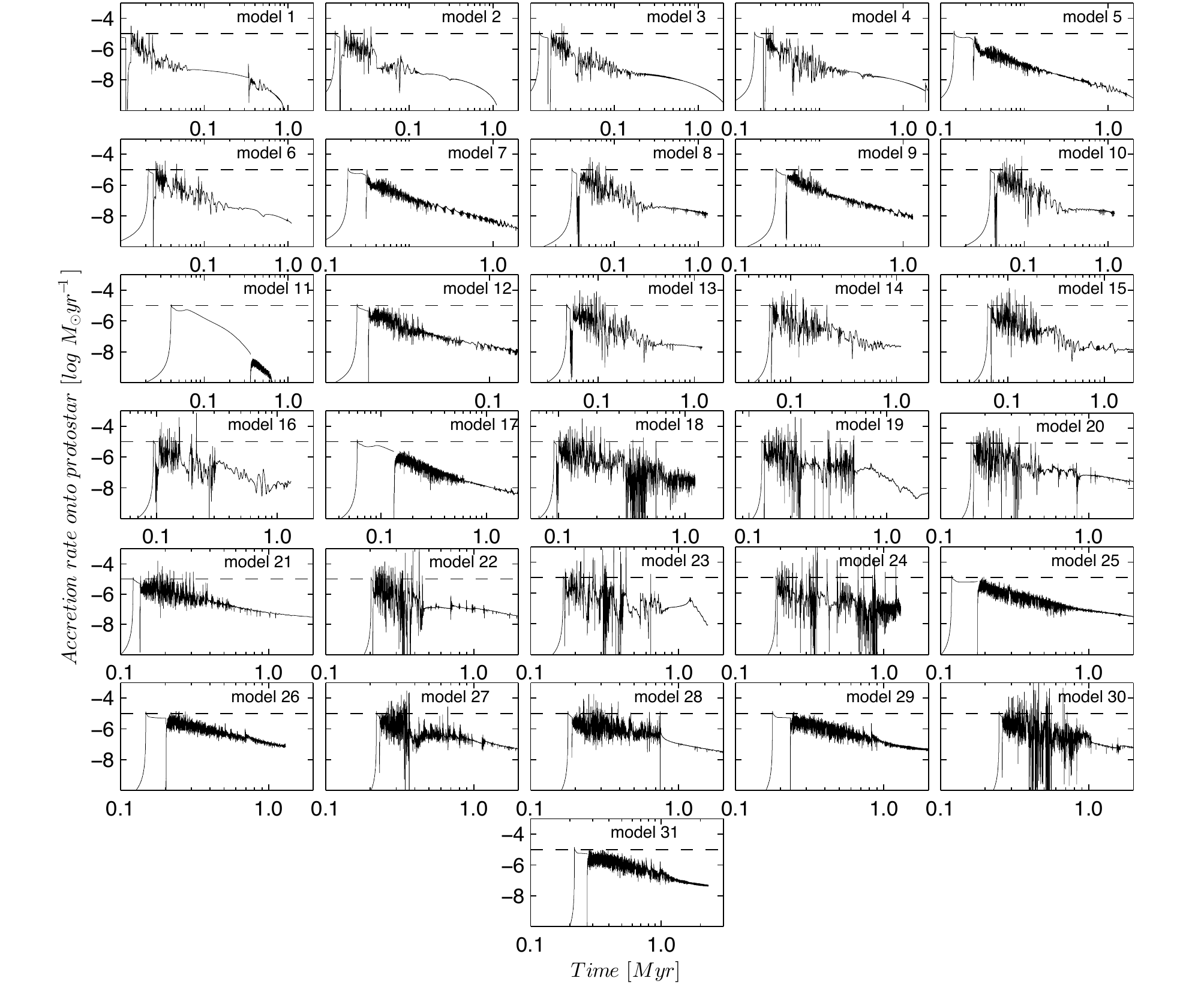}
  \caption{Mass accretion rates vs. time for the 31 models calculated with hybrid accretion. The horizontal dashed lines 
  mark the critical value for the transition from cold to hot  accretion.}
  \label{figtmdothybrid}
\end{figure*}

We provide the details of the initial parameters of pre-stellar cores in cold accretion (Table~\ref{table1}) and in the hybrid accretion simulations (Table~\ref{table2}).
In both tables, the second column is the initial core mass, third column
is the ratio of rotational to gravitational energy, fourth column is the central angular velocity, fifth
column is the radius of the central near-constant-density plateau, and sixth column is the central surface
density. In addition, the seventh column provides the final stellar masses in each model. The evolution with time of the mass accretion rates onto the protostar since the beginning of the collapse of the pre-stellar cores for all simulations are shown in  Figures \ref{figtmdotcold} and \ref{figtmdothybrid}. 

\begin{table*}[h]
\protect\caption{Cold accretion models}
\label{table1}
\centering{}%
\begin{tabular}{ccccccc}
\hline 
Model  & $M_{\mathrm{core}}$  & $\beta$  & $\Omega_{0}$  & $r_{\mathrm{0}}$  & $\Sigma_{0}$  & $M_{\ast,{\rm fin}}$ \tabularnewline
 & $M_{\odot}$  & $\%$  & km~s$^{-1}$~pc$^{-1}$  & AU  & g~cm$^{-2}$  & $M_{\odot}$ \tabularnewline
\hline 
1  & 0.06  & 2.87  & 45.0  & 137  & 0.9  & 0.047\tabularnewline
2  & 0.077  & 2.87  & 26.0  & 171  & 0.72  & 0.052\tabularnewline
3  & 0.09  & 2.87  & 30.0  & 205  & 0.6  & 0.060\tabularnewline
4  & 0.09  & 1.28  & 20.0  & 205  & 0.6  & 0.062\tabularnewline
5  & 0.092  & 0.57  & 13.3  & 205  & 0.6  & 0.072\tabularnewline
6  & 0.12  & 1.28  & 15.0  & 274  & 0.45  & 0.091\tabularnewline
7  & 0.12  & 0.57  & 10.0  & 274  & 0.45  & 0.095\tabularnewline
8  & 0.15  & 2.87  & 18.0  & 342  & 0.36  & 0.102\tabularnewline
9  & 0.15  & 1.28  & 12.0  & 342  & 0.36  & 0.112\tabularnewline
10  & 0.15  & 0.57  & 8.0  & 342  & 0.36  & 0.118\tabularnewline
11  & 0.15  & 0.14  & 4.0  & 342  & 0.36  & 0.138\tabularnewline
12  & 0.29  & 1.28  & 6.0  & 651  & 0.19  & 0.207\tabularnewline
13  & 0.31  & 0.14  & 2.0  & 685  & 0.18  & 0.215\tabularnewline
14  & 0.31  & 0.57  & 4.0  & 685  & 0.18  & 0.233\tabularnewline
15  & 0.46  & 2.87  & 6.0  & 1028  & 0.12  & 0.279\tabularnewline
16  & 0.46  & 1.28  & 4.0  & 1028  & 0.12  & 0.281\tabularnewline
17  & 0.46  & 0.57  & 2.66  & 1028  & 0.12  & 0.304\tabularnewline
18  & 0.61  & 2.87  & 4.5  & 1371  & 0.09  & 0.313\tabularnewline
19  & 0.61  & 1.28  & 3.0  & 1371  & 0.09  & 0.373\tabularnewline
20  & 0.46  & 0.14  & 1.3  & 1028  & 0.12  & 0.402\tabularnewline
21  & 0.61  & 0.57  & 2.0  & 1371  & 0.09  & 0.431\tabularnewline
22  & 0.76  & 1.28  & 2.4  & 1714  & 0.07  & 0.448\tabularnewline
23  & 0.61  & 0.14  & 1.0  & 1371  & 0.09  & 0.470\tabularnewline
24  & 0.77  & 0.57  & 1.6  & 1714  & 0.072  & 0.490\tabularnewline
25  & 0.92  & 2.87  & 3.0  & 2057  & 0.06  & 0.530\tabularnewline
26  & 1.08  & 1.28  & 1.71  & 2400  & 0.05  & 0.533\tabularnewline
27 & 1.08  & 0.57  & 1.14  & 2400  & 0.05  & 0.662\tabularnewline
28 & 1.23  & 2.87  & 2.25  & 2743  & 0.045  & 0.705\tabularnewline
29 & 0.92  & 0.14  & 0.7  & 2057  & 0.06  & 0.735\tabularnewline
30 & 1.38  & 1.28  & 1.33  & 3086  & 0.04  & 0.752\tabularnewline
31 & 1.7  & 0.57  & 0.73  & 3806  & 0.033  & 0.805\tabularnewline
32 & 1.38  & 0.57  & 0.88  & 3086  & 0.04  & 0.912\tabularnewline
33 & 1.69  & 1.28  & 1.09  & 3771  & 0.031  & 0.926\tabularnewline
34 & 1.38  & 0.14  & 0.44  & 3086  & 0.04  & 1.032\tabularnewline
35 & 1.69  & 0.14  & 0.36  & 3771  & 0.033  & 1.255\tabularnewline
36 & 2.0  & 0.14  & 0.31  & 4457  & 0.028  & 1.443\tabularnewline
\hline 
\end{tabular}
\end{table*}

\begin{table*}
\protect\caption{Hybrid accretion models}
\label{table2}
\centering{}%
\begin{tabular}{ccccccc}
\hline 
Model  & $M_{\mathrm{core}}$  & $\beta$  & $\Omega_{0}$  & $r_{\mathrm{0}}$  & $\Sigma_{0}$  & $M_{\ast,{\rm fin}}$ \tabularnewline
 & $M_{\odot}$  & $\%$  & km~s$^{-1}$~pc$^{-1}$  & AU  & g~cm$^{-2}$  & $M_{\odot}$ \tabularnewline
\hline 
1  & 0.06  & 2.87  & 45.0  & 137  & 0.9  & 0.046\tabularnewline
2  & 0.09  & 2.87  & 30.0  & 205  & 0.6  & 0.061\tabularnewline
3  & 0.09  & 1.74  & 23.3  & 205  & 0.6  & 0.070\tabularnewline
4  & 0.12  & 1.74  & 17.5  & 274  & 0.45  & 0.091\tabularnewline
5  & 0.11  & 0.57  & 11.4  & 240  & 0.52  & 0.092\tabularnewline
6  & 0.15  & 2.87  & 18.0  & 342  & 0.36  & 0.101\tabularnewline
7  & 0.14  & 0.57  & 8.88  & 308  & 0.4  & 0.118\tabularnewline
8  & 0.23  & 1.74  & 9.33  & 514  & 0.24  & 0.166\tabularnewline
9  & 0.23  & 0.57  & 5.33  & 514  & 0.24  & 0.178\tabularnewline
10  & 0.29  & 1.74  & 7.0  & 651  & 0.19  & 0.202\tabularnewline
11  & 0.31  & 0.14  & 2.0  & 685  & 0.18  & 0.215\tabularnewline
12  & 0.31  & 0.57  & 4.0  & 685  & 0.18  & 0.236\tabularnewline
13  & 0.35  & 1.74  & 6.1  & 788  & 0.16  & 0.237\tabularnewline
14  & 0.46  & 2.87  & 6.0  & 1028  & 0.12  & 0.263\tabularnewline
15  & 0.46  & 1.74  & 4.66  & 1028  & 0.12  & 0.288\tabularnewline
16  & 0.69  & 2.87  & 4.0  & 1543  & 0.08  & 0.348\tabularnewline
17  & 0.46  & 0.14  & 1.3  & 1028  & 0.12  & 0.402\tabularnewline
18  & 0.69  & 1.74  & 3.11  & 1543  & 0.08  & 0.459\tabularnewline
19  & 0.94  & 1.74  & 2.33  & 2091  & 0.06  & 0.534\tabularnewline
20  & 1.21  & 1.74  & 1.75  & 2708  & 0.046  & 0.556\tabularnewline
21  & 0.92  & 0.57  & 1.33  & 2057  & 0.06  & 0.614\tabularnewline
22  & 1.53  & 1.74  & 1.4  & 3428  & 0.036  & 0.708\tabularnewline
23  & 1.23  & 2.87  & 2.25  & 2743  & 0.045  & 0.709\tabularnewline
24  & 1.38  & 2.87  & 2.1  & 3086  & 0.04  & 0.715\tabularnewline
25  & 0.92  & 0.14  & 0.7  & 2057  & 0.06  & 0.735\tabularnewline
26  & 1.15  & 0.14  & 0.53  & 2571  & 0.048  & 0.832\tabularnewline
27 & 1.7  & 0.57  & 0.73  & 3806  & 0.033  & 0.884\tabularnewline
28 & 1.38  & 0.57  & 0.88  & 3086  & 0.04  & 0.923\tabularnewline
29 & 1.38  & 0.14  & 0.44  & 3086  & 0.04  & 1.016\tabularnewline
30 & 1.92  & 0.57  & 0.64  & 4286  & 0.029  & 1.072\tabularnewline
31 & 1.69  & 0.14  & 0.36  & 3771  & 0.033  & 1.234\tabularnewline
\hline 
\end{tabular}
\end{table*}

\bibliographystyle{aa.bst}
\bibliography{references}

\end{document}